\documentclass[a4paper]{revtex4}
\usepackage{graphicx}
\usepackage{fancyhdr}
\usepackage{amsmath}
\usepackage{color}

\begin{document}

\title{XMASS: Recent results and status}

%

\author{K. Hiraide$^{1,2}$, for the XMASS Collaboration}
\affiliation{$^1$Kamioka Observatory, Institute for Cosmic Ray Research, the University of Tokyo,
         Higashi-Mozumi, Kamioka, Hida, Gifu, 506-1205, Japan\\
         $^2$Kavli Institute for the Physics and Mathematics of the Universe, the University of Tokyo,
         Kashiwa, Chiba, 277-8582, Japan}

\begin{abstract}
The XMASS project is designed for multiple physics goals using highly-purified
liquid xenon scintillator in an ultra-low radioactivity environment.
As the first stage of the project, the detector with 835~kg of liquid xenon
was constructed and is being operated.
In this paper, we present results from our commissioning data,
current status of the experiment, and a next step of the project.
\end{abstract}

\maketitle

\thispagestyle{fancy}


\section{Introduction}
The XMASS project is designed to detect dark matter, neutrinoless double beta decay,
and $^{7}$Be/$pp$ solar neutrinos using highly-purified liquid xenon scintillator
in an ultra-low radioactivity environment \cite{xmass-suzuki}.
The advantages of using liquid xenon are a large amount of scintillation light yield,
scalability of the size of the detector mass, an easy purification to reduce internal radioactive backgrounds,
and a high atomic number ($Z=54$) to shield radiations from outside of the detector.
As the first stage of the XMASS project (XMASS-I), the detector with 835~kg of liquid xenon was constructed.
Its construction started in April 2007 and completed in September 2010.
After completion of the detector, commissioning data was taken from December 2010 to May 2012.
In order to reduce the backgrounds, detector refurbishment was conducted.
After a year of the detector refurbishment, data-taking resumed in November 2013
and is continuing for more than a year till now.

In this paper, we present physics results from our commissioning data,
the detector refurbishment and current status of the experiment,
and a next step of the project.

\section{The XMASS-I detector}
XMASS-I is a single phase liquid xenon scintillator detector located underground
(2700~m water equivalent) at the Kamioka Observatory.
It contains 835~kg of liquid xenon in an active region.
The volume is viewed by 630 hexagonal and 12 cylindrical Hamamatsu R10789 photomultiplier tubes (PMTs)
arranged on an 80 cm diameter pentakis-dodecahedron support structure.
A total photocathode coverage of more than 62\% is achieved.
The spherical arrays of PMTs are arranged in a double wall vessel
made of oxygen free high conductivity (OFHC) copper.
The detector is calibrated regularly with a $^{57}$Co source inserted along the
central vertical axis of the detector.
By the data taken with the $^{57}$Co source at the center of the detector volume,
the photoelectron yield was determined to be $\sim$14~photoelectrons/keV$_{ee}$
where the subscript $ee$ stands for the electron equivalent energy deposition.
In order to shield the liquid xenon detector from external gammas, neutrons,
and muon-induced backgrounds, the copper vessel was placed at the center of
a $\phi$10~m$\times$ 10.5~m cylindrical tank filled with pure water.
The water tank is equipped with 72 Hamamatsu R3600 20-inch PMTs
to provide both an active muon veto and passive shielding against these backgrounds.
XMASS-I is the first direct detection dark matter experiment equipped with such
an active water Cherenkov shield.
The liquid xenon and water Cherenkov detectors are hence called an Inner Detector (ID) and
an Outer Detector (OD), respectively.
More details are described in Ref.\ \cite{xmass-detector}.

\section{Results from 6.7 live days of the low energy threshold data}
Owing to the large photoelectron yield we achieved, the XMASS-I detector has an advantage
in lowering the energy threshold.
Hence, a part of commissioning data was taken with a low trigger threshold of
four PMT hits which corresponds to 0.3~keV$_{ee}$.
Two physics results were obtained using 6.7 live days of data collected with
the lowest energy threshold.
In order to achieve optimal sensitivity, the entire detector mass of 835~kg was used
because fiducialization is increasingly difficult at these low energies.

\subsection{Search for light WIMPs~\cite{xmass-lowmasswimp}}
\begin{figure}[tbp]
\begin{center}
  \begin{tabular}{cc}
    \begin{minipage}{80mm}
      \includegraphics[height=7cm]{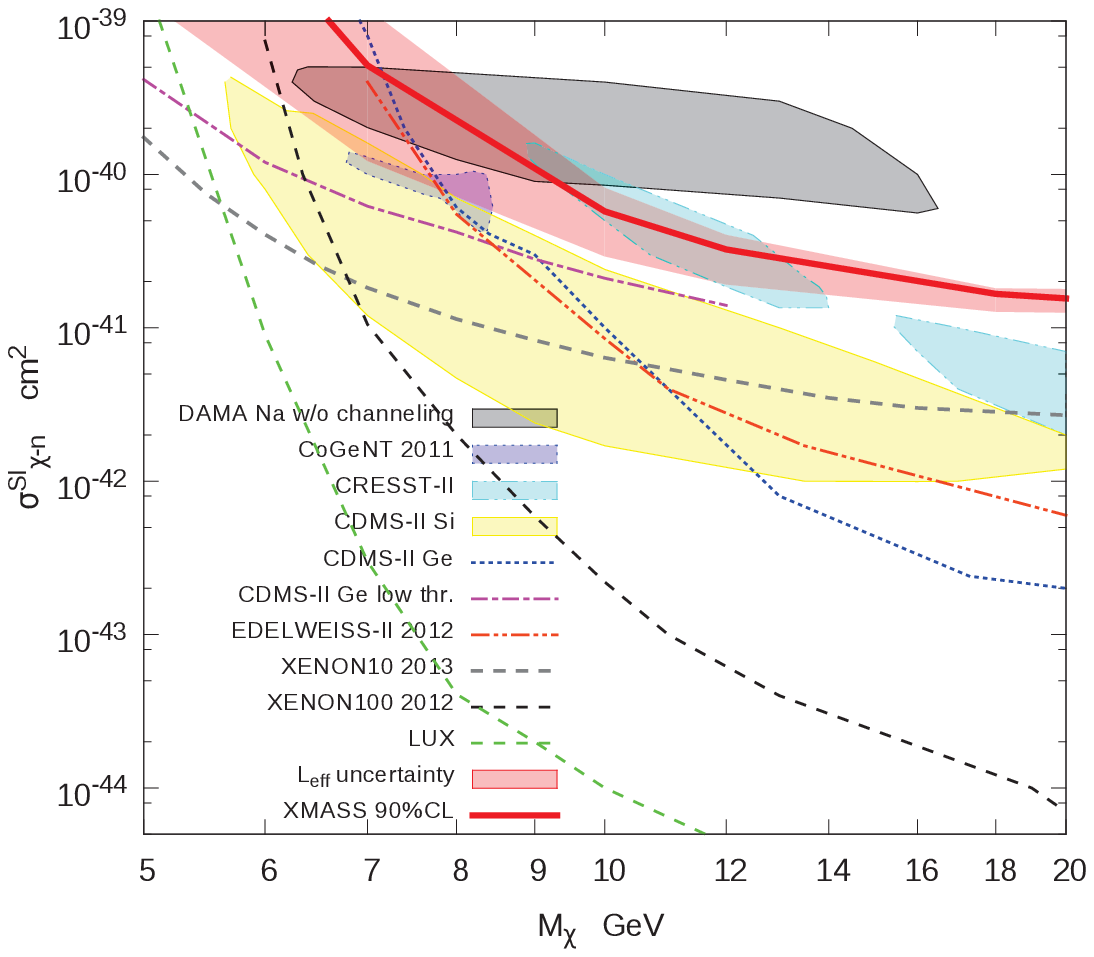}
      \caption{Obtained 90\% C.L. limits on spin-independent WIMP-nucleon cross section as a function of WIMP mass.}
      \label{xmass:lowmasswimp}
    \end{minipage}
    &
    \begin{minipage}{80mm}
      \includegraphics[height=7cm]{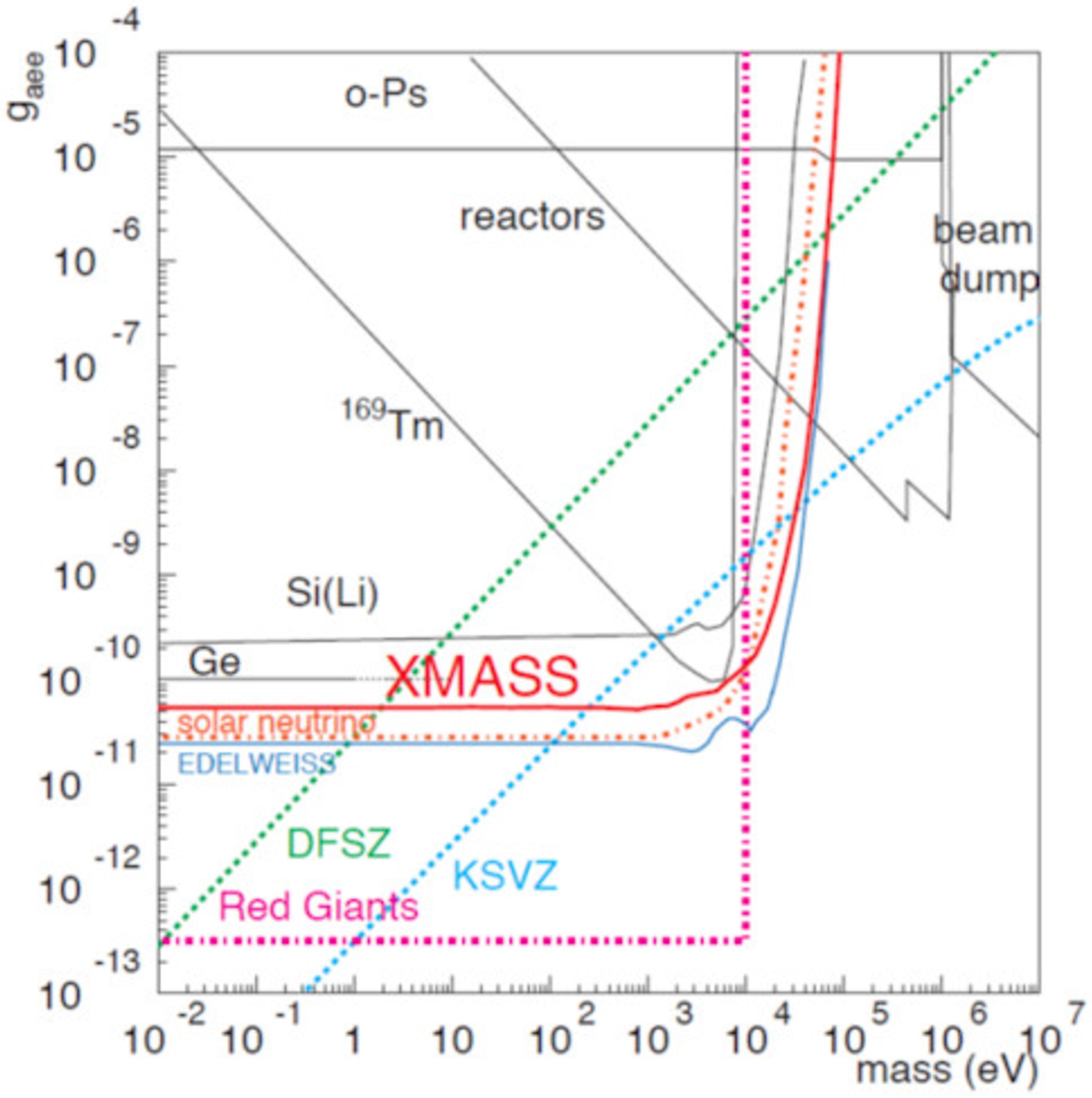}
      \caption{Obtained 90\% C.L. limits on the axion-electron coupling constant ($g_{aee}$) as a function of axion mass.}
      \label{xmass:solaraxion}    
    \end{minipage}
  \end{tabular}
\end{center}
\end{figure}

Weakly Interacting Massive Particles (WIMPs), the most possible dark matter candidates,
can be detected directly through observation of nuclear recoils produced in
their elastic scattering interactions with detector nuclei.
Although many theories of physics beyond the Standard Model predict WIMPs with mass larger than 100~GeV,
some experiments indicate a possible WIMP signal with a lighter mass of
$\sim$10~GeV~\cite{Bernabei:2008yi,Aalseth:2010vx,Angloher:2011uu}.
We performed a search for low mass WIMPs using the 6.7 live days of low energy threshold data.
The observed spectrum does not have any prominent feature which suggests positive evidence of
WIMP signals over background.
Figure~\ref{xmass:lowmasswimp} shows the resulting 90\% confidence level (C.L.) limit.
The impact of the uncertainty from the scintillation efficiency,
$\mathcal{L}_{\rm eff}$, is shown separately in the figure.
Without discriminating between nuclear-recoil and electronic events,
XMASS sets an upper limit on the WIMP-nucleon cross section for WIMPs with
masses below 20~GeV and excludes part of the parameter space allowed by other experiments.

\subsection{Search for solar axion~\cite{xmass-solaraxion}}
Axion is a hypothetical particle which is invented for solving the CP problem in the strong interactions.
The particle would be produced in the sun through various mechanisms;
in this paper, we focus on Compton scattering of photons on electrons, $e + \gamma \to e + a$,
and bremsstrahlung of axions from electrons, $e + Z \to e + a + Z$~\cite{Derbin:2012yk}.
In the liquid xenon detector, axions can be detected through the axio-electric effect,
which is an analog of the photo-electric effect.
In order to search for solar axion in XMASS, the same data set as used for light WIMPs search is analyzed.
No prominent feature which suggests a positive evidence of axion signals over background is observed.
Hence, by adopting a criteria that the expected signal cannot be larger than the observed spectrum
at 90\% C.L., upper limits on the axion-electron coupling constant, $g_{aee}$, are derived.
Figure~\ref{xmass:solaraxion} shows a summary of the upper limits on $g_{aee}$.
The best direct experimental limit on $g_{aee}$ is obtained, and the limit is
close to the one obtained theoretically based on the consistency between the observed and expected
solar neutrino fluxes. 

\section{Results from the full 165.9 live days of commissioning data}
In our recent analyses, we have achieved an unprecedented low-background level of
$\sim 10^{-4}$\,${\rm day}^{-1}{\rm kg}^{-1}{\rm keV}_{ee}^{-1}$ in the energy range
around a few tens of keV$_{ee}$. This background level is an order of magnitude
lower than XENON100 and LUX before their $e/\gamma$ rejection.
Three physics results were obtained from our 165.9 live days of the commissioning data
with a restricted target mass of 41~kg at the central region of the detector.

\subsection{Search for inelastic WIMP-nucleus scattering on $^{129}$Xe \cite{xmass-inelastic}}
\begin{figure}[tbp]
\begin{center}
  \begin{tabular}{cc}
    \begin{minipage}{80mm}
      \includegraphics[width=8cm]{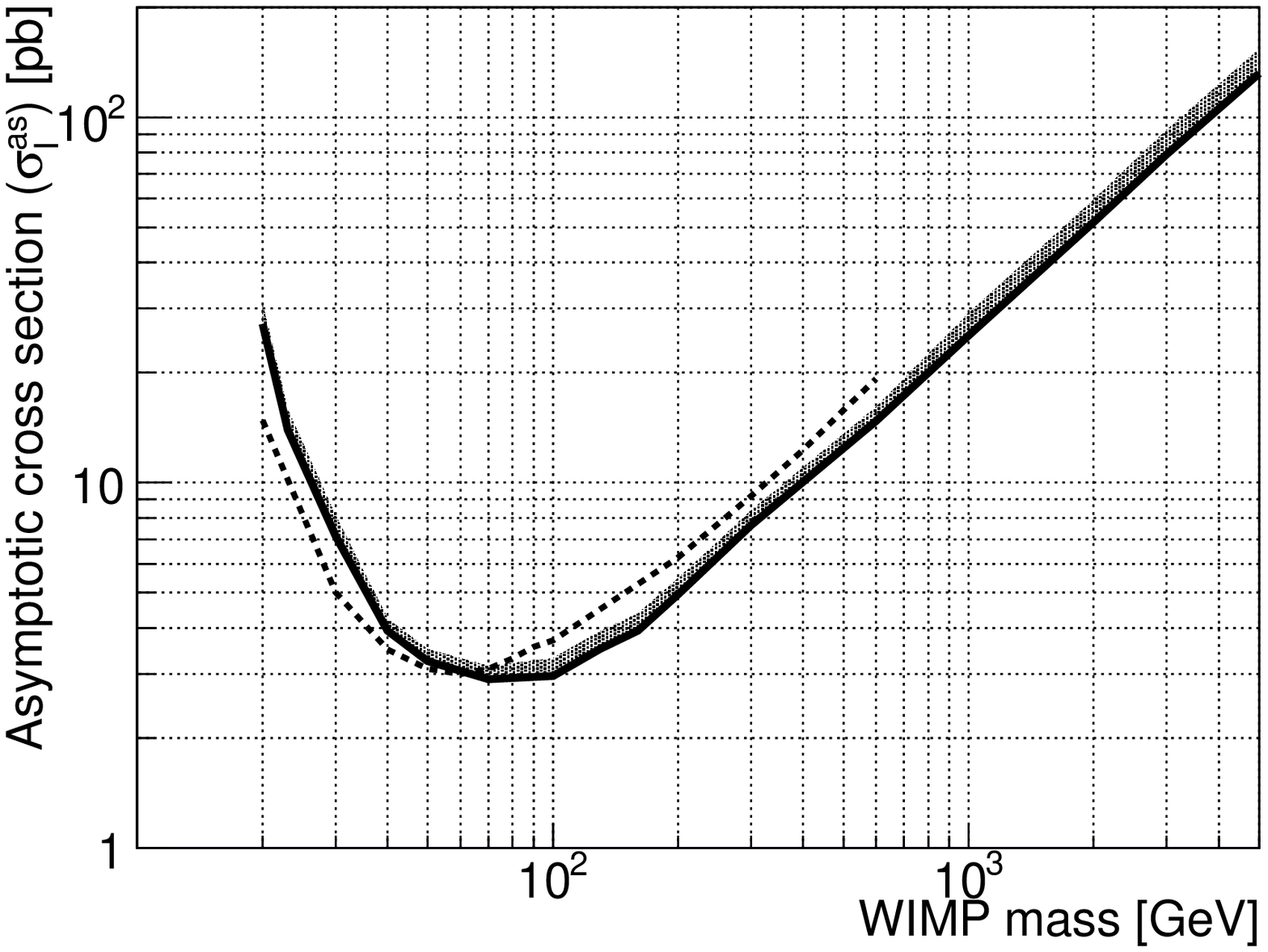}
      \vspace{3mm}
      \caption{The solid line is our 90\% C.L. upper limit on the asymptotic cross section
      for inelastic scattering on $^{129}$Xe.
      The gray band covers its variation with our systematic uncertainty.
      The dashed line is the limit obtained by the DAMA group.}
      \label{xmass:inelastic}  
    \end{minipage}
    &
    \begin{minipage}{80mm}
      \includegraphics[width=8cm]{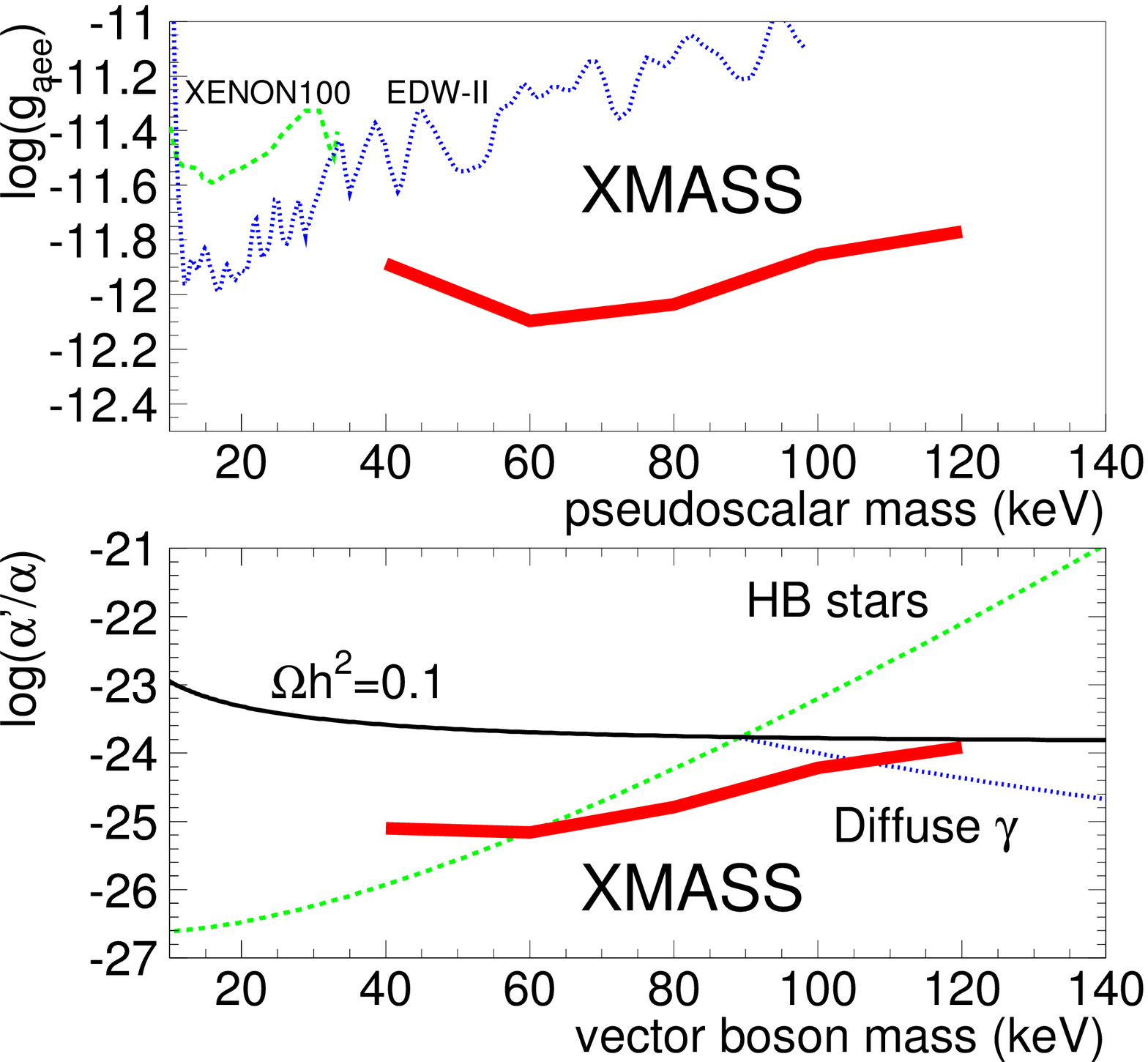}
      \caption{Obtained 90\% C.L. upper limits on coupling constants for
      electrons and pseudoscalar super-WIMPs (top) and for electrons and vector super-WIMPs (bottom).}
      \label{xmass:superwimps}
   \end{minipage}      
  \end{tabular}
\end{center}
\end{figure}

Inelastic scattering that excites low-lying nuclear states in suitable target nuclei
provides another avenue to probe WIMP dark matter.
Its advantage is that nuclear excited states and their de-excitation mechanisms are
typically well known, and thus the expected energy deposit in the detector is calculated,
resulting in the readily identifiable signature of a line in the energy spectrum.
$^{129}$Xe has the lowest-lying excited nuclear state at 39.58 keV and
almost the highest natural abundance of 26.4\% among the xenon isotopes.
A high-energy tail of the peak at 39.58~keV is expected since the scintillation lights from
the recoil of $^{129}$Xe nucleus and from the nuclear de-excitation $\gamma$-ray
cannot be separated since lifetime of the excited state is much shorter
than the decay time constant of scintillation light.
We searched for such an inelastic scattering of WIMPs on $^{129}$Xe.
No significant excess of the events was observed, and hence the 90\% C.L.
upper limits on the cross section for inelastic scattering on $^{129}$Xe was derived
as shown in Fig.~\ref{xmass:inelastic}.

\subsection{Search for bosonic superweakly interacting massive dark matter particles \cite{xmass-superwimps}}
Although a WIMP dark matter is a well-motivated model and fits the cold dark matter paradigm,
simulations based on this cold dark matter scenario expect a richer structure on galactic scales than those observed.
Furthermore, there is so far no evidence of supersymmetric particles at the LHC, and therefore,
it is important to investigate various types of dark matter candidates~\cite{Pospelov:2008jk,Redondo:2008ec}.
These facts strengthen an interest to consider lighter and more weakly interacting particles
such as super-WIMPs, a warm dark matter candidate. 
Bosonic super-WIMPs are experimentally interesting since their absorption in a target material
would deposit an energy essentially equivalent to the super-WIMP's rest mass.
We conducted a search for vector and pseudoscalar super-WIMPs in the mass range between 40 and 120 keV.
No significant excess above background was observed, and hence the 90\% C.L. upper limits
on coupling constants for pseudoscalar bosons and vector bosons were obtained as shown in
Fig.~\ref{xmass:superwimps}.
This is the first direct detection experiment exploring the vector super-WIMPs and
the obtained limit for the vector super-WIMPs excludes the possibility that
such particles constitute all of the dark matter.
The absence of the signal also provides the most stringent direct constraint
on the coupling constant of pseudoscalar dark matter to electrons.
This result was published in Physical Review Letters, selected as Editors' Suggestion.

\subsection{Search for two-neutrino double electron capture on $^{124}$Xe}
\begin{figure}[h]
\begin{center}
  \begin{tabular}{cc}
    \begin{minipage}{80mm}
    \includegraphics[width=8cm]{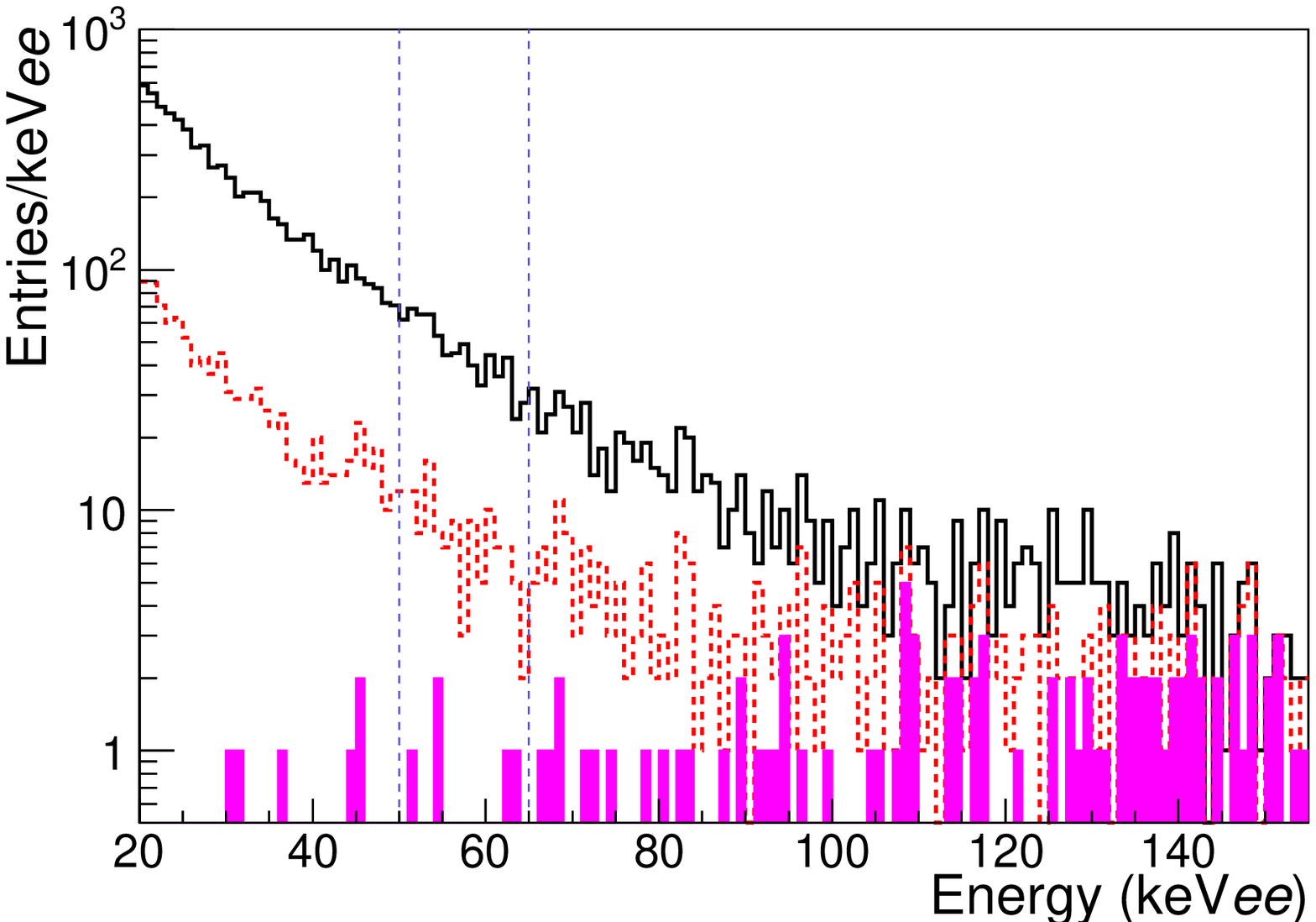}
    \end{minipage}
    &
    \begin{minipage}{80mm}
    \includegraphics[width=8cm]{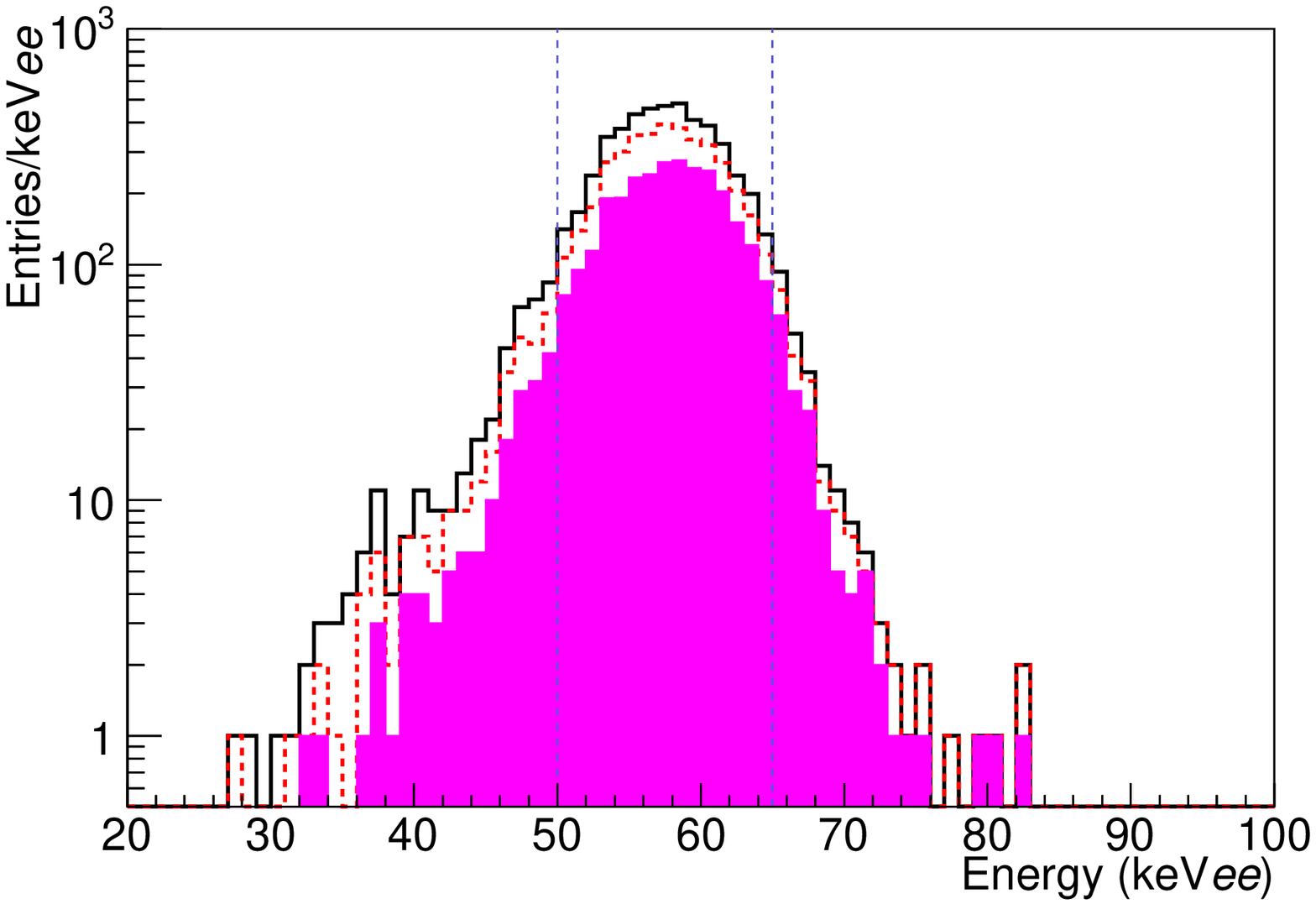}
   \end{minipage}      
  \end{tabular}
  \caption{Energy spectra of the observed events (left) and simulated events (right)
  after each reduction step for two-neutrino double electron capture on $^{124}$Xe.
  Filled histograms show remaining events after applying radius cut,
  timing cut and hit pattern cut. Vertical dashed lines indicate the energy window for signal.}
  \label{xmass:2nuecec}  
\end{center}
\end{figure}

Neutrinoless double beta decay and its inverse, neutrinoless double electron capture,
are lepton number violating processes, and their existence is an evidence
that neutrino is a Majorana particle.
On the other hand, two-neutrino modes of double beta decay and double electron capture
are allowed within the standard model of particle physics.
Although two-neutrino double beta decay has been observed in more than ten isotopes,
there exists only a few positive experimental results for two-neutrino double electron
capture so far: a geochemical measurement for $^{130}$Ba and a direct measurement for $^{78}$Kr.
Any measurement of two-neutrino double electron capture will provide a new reference
for the calculation of nuclear matrix elements.
Natural xenon contains the double electron capture nuclei $^{124}$Xe (0.095\%).
In the case that two $K$-shell electron in the $^{124}$Xe atom are captured simultaneously,
the daughter atom of $^{124}$Te is formed with two vacancies in the $K$-shell
and de-excites by by emitting atomic $X$-rays and/or Auger electrons.
The total energy deposition in the detector is 63.6~keV, which is twice of the binding
energy of a $K$-shell electron in a tellurium atom.
We performed a search for two-neutrino double electron capture on $^{124}$Xe.
Fig.~\ref{xmass:2nuecec} shows energy spectra of the observed events and simulated events
after each reduction step.
After all cuts, 5 events remain in the signal region while the expected background is
$5.3\pm 0.5$ events originating from the $^{222}$Rn daughter $^{214}$Pb in the detector.
No significant peak above background was observed and hence we set the 90\% C.L. lower limit
on the half-life of $4.9\times 10^{21}$~years after subtracting background,
which provides the most stringent limit for $^{124}$Xe.

\section{Detector refurbishment and current status}
\begin{figure}[tbp]
\begin{center}
  \begin{tabular}{cc}
    \begin{minipage}{80mm}
    \includegraphics[width=8cm]{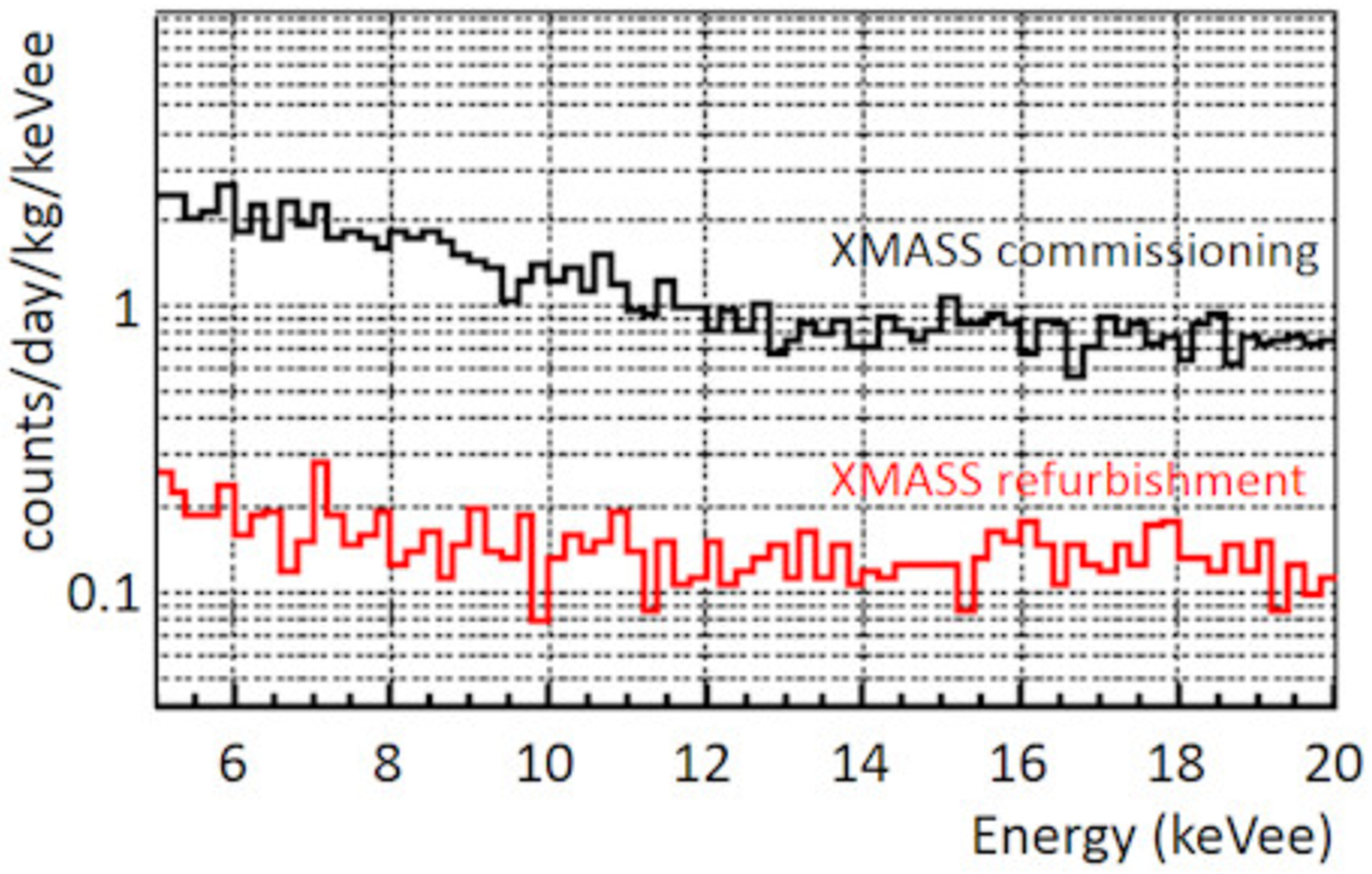}
    \end{minipage}
    &
    \begin{minipage}{80mm}
    \includegraphics[width=8cm]{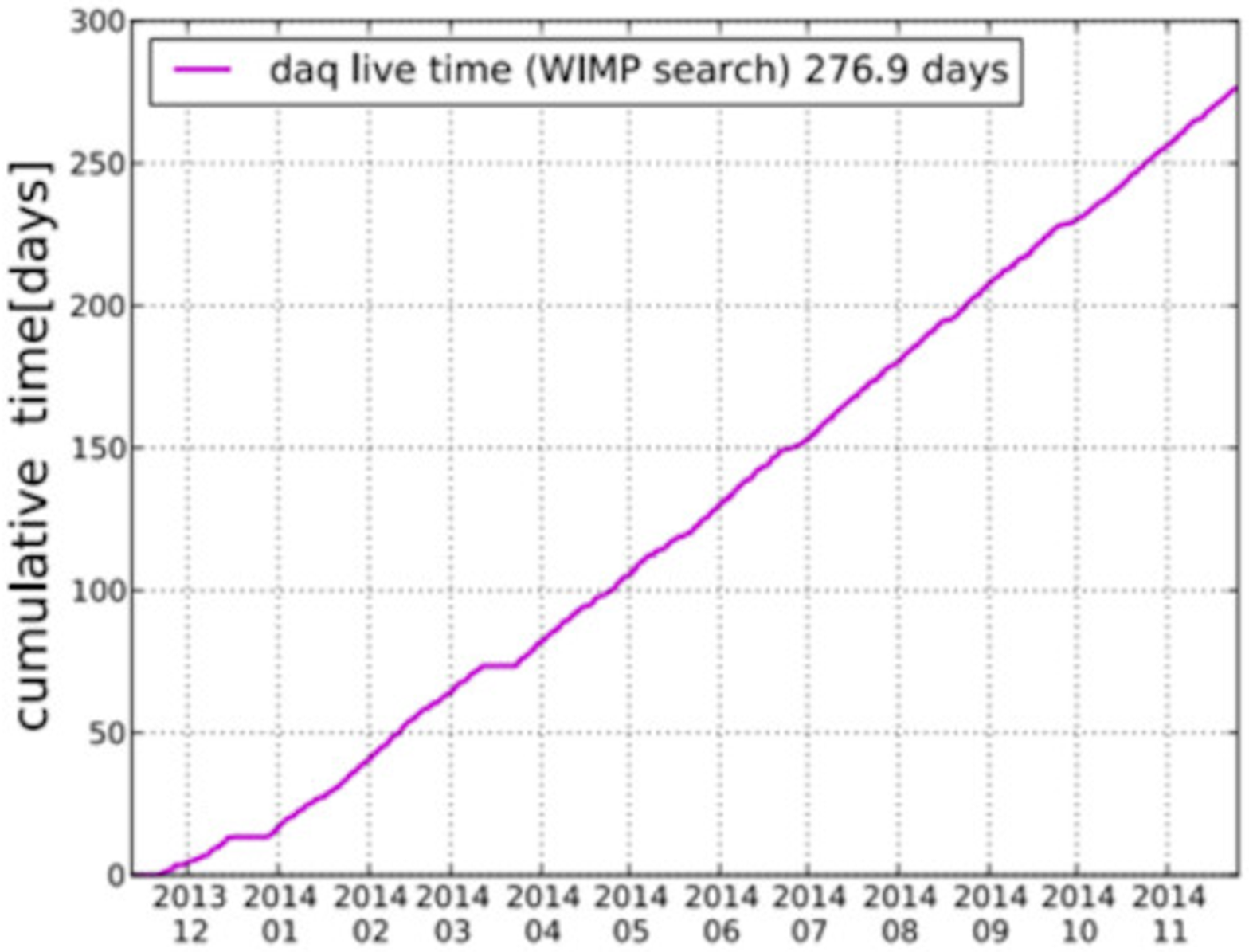}
   \end{minipage}      
  \end{tabular}
  \caption{(Left) Energy spectra for the entire detector volume before and after the detector refurbishment.
  One order of magnitude reduction of the event rate is achieved.
  (Right) History of the accumulated livetime after data quality cuts after the refurbishment.}
  \label{xmass:refurbishment}  
\end{center}
\end{figure}

During the commissioning data-taking, we found that a majority of events at low energy
originated from radioactive contamination in the aluminum seal of the PMT window.
The contaminated parts of PMTs were covered by copper rings and plates in order to
stop scintillation lights and radiations caused by its contamination.
PMTs were cleaned by acid and copper parts were electropolished
in order to remove possible surface contamination.
After completion of the detector refurbishment in November 2013,
data-taking resumed and is continuing till now.
Fig.~\ref{xmass:refurbishment}~(left) shows energy spectra for the entire
detector volume between before and after the detector refurbishment.
One order of magnitude reduction of the event rate in this energy range was achieved.
Fig.~\ref{xmass:refurbishment}~(right) shows a history of the accumulated livetime
after data quality cuts since November 2013.
One of the physics targets using these data is a dark matter search by looking for
an annual modulation of the event rate.
The count rate of dark matter signal would modulate with a period of a year due to
the relative motion of the Earth around the Sun.
The dark matter flux at the detector becomes maximal in June and minimal in December.
The annual modulation is a strong signature for dark matter since
most of backgrounds are not expected to show this kind of time dependence.
XMASS has a sensitivity to the annual modulation claimed by the DAMA/LIBRA group
without particle identification. These analyses are under way.

\section{Next step: XMASS-1.5}
\begin{figure}[h]
\begin{center}
  \begin{tabular}{cc}
    \begin{minipage}{60mm}
    \includegraphics[width=6cm]{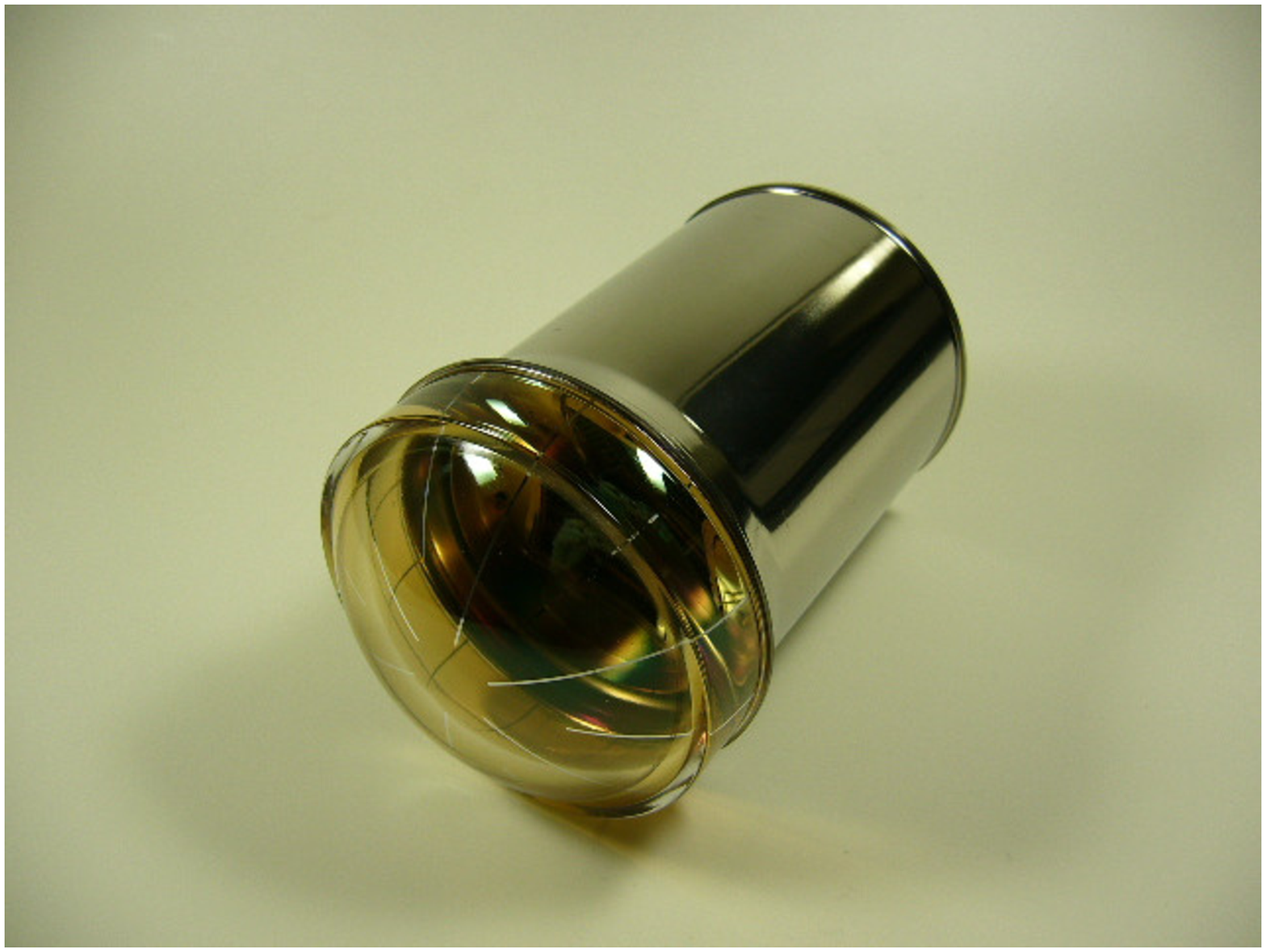}
    \end{minipage}
    &
    \begin{minipage}{100mm}
    \includegraphics[width=10cm]{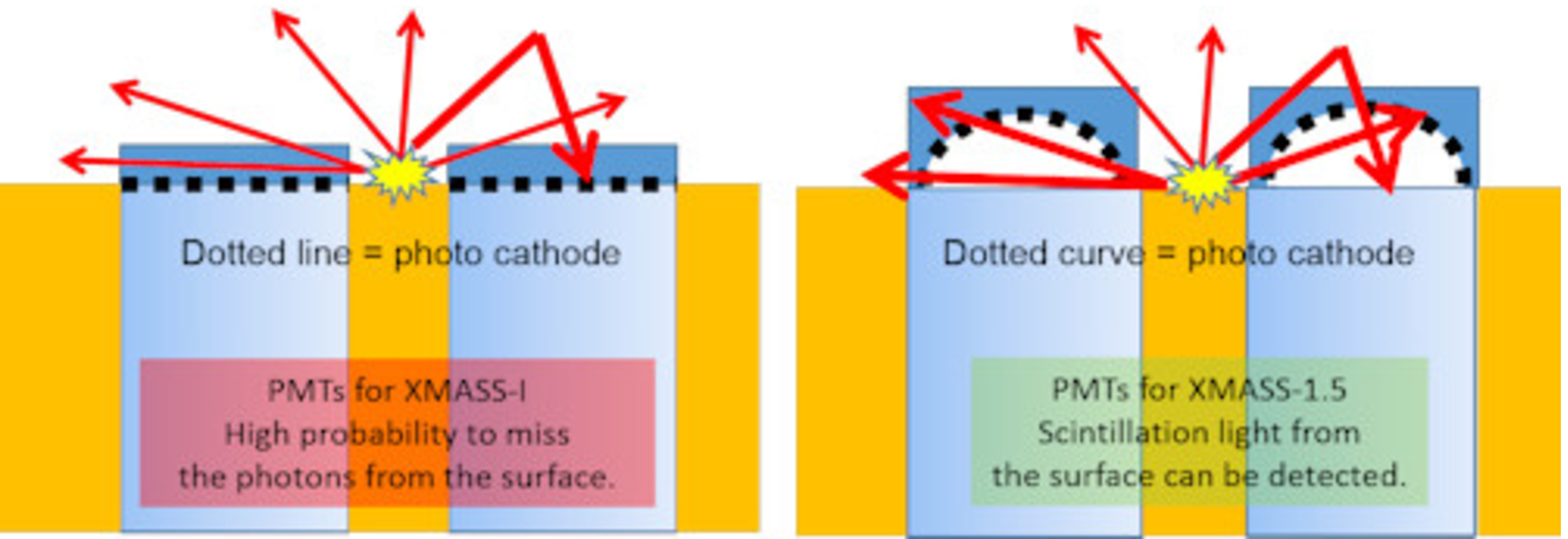}
   \end{minipage}      
  \end{tabular}
  \caption{(Left) Picture of a newly-developed 3-inch PMT for the next generation XMASS detector.
  (Right) Comparison of response to surface events of PMT with a flat photocathode and 
  a dome-shaped photocathode. Since the dome-shaped photocathode has a large acceptance
  for scintillation light from the inner surface of the detector, identification 
  of surface events is much more efficient with the dome-shaped PMTs.}
  \label{xmass:newpmt}  
\end{center}
\end{figure}

\begin{figure}[h]
  \begin{center}
    \includegraphics[width=12cm]{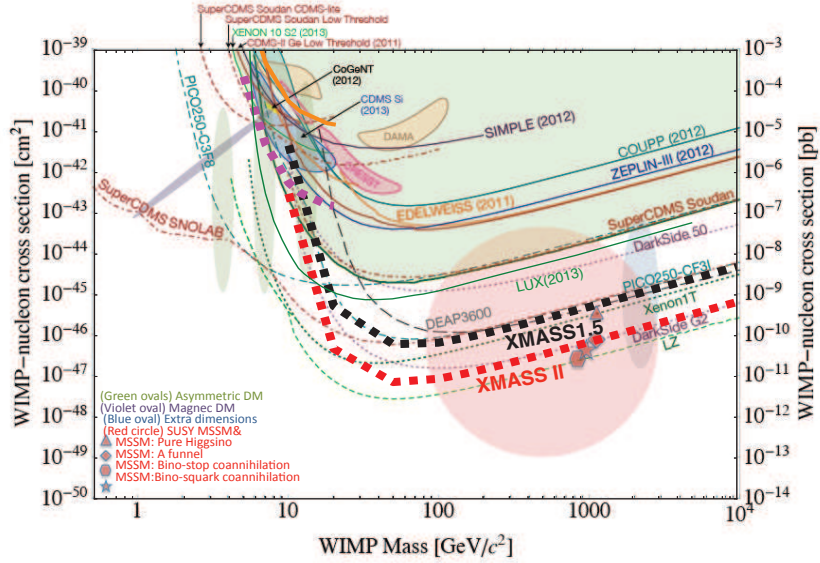}
  \end{center}
  \caption{Expected sensitivity with XMASS-1.5 (XMASS-II)
  assuming fiducial mass of 1~ton (10~tons) and 1-year (5-year) running, respectively.}
  \label{xmass:sensitivity}
\end{figure}

We are planning to build the next stage detector, XMASS-1.5, with total 5 tons of liquid xenon.
The basic design of the detector is same as the one for the XMASS-I detector
but with an essential improvement of the discrimination between the surface background
and the inner events.
For this purpose, we have developed a new type of PMT which has a dome-shaped photocathode
as shown in Fig.~\ref{xmass:newpmt} (left).
Fig.\ \ref{xmass:newpmt} (right) illustrates a comparison of response to surface events
of PMT with a flat photocathode and a dome-shaped photocathode.
The dome-shaped PMTs are more efficient for detecting scintillation photons originated
at the inner surface of the detector than the flat-shaped PMTs used for the XMASS-I detector.
Fig.~\ref{xmass:sensitivity} shows the expected sensitivity to WIMP-nucleon
cross section with the XMASS-1.5 detector.
With newly-developed PMTs, we can search for heavy WIMPs with cross sections
less than 10$^{-46}$~cm$^{2}$ even for the same background level of the XMASS-I.
Beyond XMASS-1.5, XMASS-II with total 24 tons of liquid xenon is planned.
The expected sensitivity to WIMP-nucleon cross section with the XMASS-II detector
is also shown in Fig.~\ref{xmass:sensitivity}.

\section{Conclusions}
The XMASS project is designed for multiple physics goals using liquid xenon.
The XMASS-I detector achieved a low-energy threshold of 0.3~keV$_{ee}$ and 
an unprecedented low-background level of
$\sim 10^{-4}$\,${\rm day}^{-1}{\rm kg}^{-1}{\rm keV}_{ee}^{-1}$ in the energy range
around a few tens of keV$_{ee}$ without $e/\gamma$ rejection.
We obtained a wide variety of physics results from our commissioning data
taken from December 2010 to May 2012.
After a year of the detector refurbishment work, data-taking resumed in November 2013
and is continuing till now. Physics results from data taken after refurbishment will come soon.
As a next step of the project, we are planning to build the next generation detector,
XMASS-1.5, with total 5 tons of liquid xenon.


\bigskip 

\end{document}